\documentclass[12pt]{article}
\usepackage{amsmath}
\def\be{\begin{equation}}
\def\ee{\end{equation}}
\def\arr{\begin{array}{rll}}
\def\ea{\end{array}}
\def\bea{\begin{eqnarray}}
\def\eea{\end{eqnarray}}

\def\N2{$N{=}2$}

\def\>{\rangle}
\def\<{\langle}
\def\+{\dagger}
\def\={\ =\ }

\begin{document}
\renewcommand{\thefootnote}{\fnsymbol{footnote}}
\begin{titlepage}
\setcounter{page}{0}
\vskip 1cm
\begin{center}
{\LARGE\bf  Ruijsenaars-Schneider three--body  }\\
\vskip 0.5cm
{\LARGE\bf models with $N=2$ supersymmetry }\\
\vskip 1cm
$
\textrm{\Large Anton Galajinsky \ }
$
\vskip 0.7cm
{\it
School of Physics, Tomsk Polytechnic University,
634050 Tomsk, Lenin Ave. 30, Russia} \\
{e-mail: galajin@tpu.ru}

\end{center}
\vskip 1cm
\begin{abstract} \noindent
The Ruijsenaars-Schneider models are conventionally regarded as
relativistic generalizations of the Calogero integrable systems.
Surprisingly enough, their supersymmetric generalizations escaped attention. In this work,
$N=2$ supersymmetric extensions of the rational and hyperbolic Ruijsenaars-Schneider three--body models are constructed
within the framework of the Hamiltonian formalism. It is also known that the rational model can be described by the geodesic equations associated with a metric connection.
We demonstrate that the hyperbolic systems are linked to non--metric connections.
\end{abstract}


\end{titlepage}

\renewcommand{\thefootnote}{\arabic{footnote}}
\setcounter{footnote}0

\noindent
{\bf 1. Introduction}\\

\noindent
The Ruijsenaars-Schneider models \cite{RS} provide interesting examples of integrable many--body systems in $d=1$ whose equations of motion involve particle velocities. They exhibit the Poincar\'e symmetries in $1+1$ dimensions, which involve translations in the temporal and spatial directions and a boost, and reduce to the Calogero systems \cite{C} in the non\-relativistic limit \cite{RS}. By this reason, the former are conventionally regarded as
the relativistic generalizations of the latter.

An important aspect of the extensive studies of the Calogero models over the last two decades has been the construction of $N=4$ supersymmetric extensions \cite{W,GLP,FIL,KL}. Interest in such systems stems from the fact that some of them are expected to be relevant for a microscopic description of the extreme black holes \cite{GT}. Worth mentioning also is that $N$--extended supersymmetry in
$d=1$ exhibits peculiar features which are absent in higher dimensions.

Surprisingly enough, supersymmetric extensions of the relativistic counterparts of the Calogero models remain almost completely unexplored.
An integrable $N=2$ supersymmetric generalization of the quantum
trigonometric Ruijsenaars-Schneider model has been reported in \cite{BDM} whose eigenfunctions  were linked to the Macdonald superpolynomials. Note, however, that the fermionic variables in \cite{BDM} and their adjoints obey  the non--standard anticommutation relations which reduce to the conventional ones in the non--relativistic limit only.

The goal of this work is to construct $N=2$ supersymmetric extensions of the rational and hyperbolic Ruijsenaars-Schneider three--body models within the framework of the Hamiltonian (on-shell) formalism. As is known, the systems admit more than one Hamiltonian description \cite{C,C1}. For a supersymmetric extension to be feasible, we suggest to choose a Hamiltonian each term of which is positive definite.

The paper is organized as follows. In subsections 2.1, 2.2, and 2.3 we briefly review the basic properties of the rational and hyperbolic Ruijsenaars-Schneider three--body models with a particular emphasis on the issue of (super)integrability. An interesting feature of these systems is that they admit an alternative description in terms of geodesic equations associated with an affine connection \cite{ABHL}. For the rational model the latter is known to be a metric connection and the manifold is actually flat \cite{ABHL}. In subsection 2.4. we demonstrate that the hyperbolic models are linked to non--metric connections. In section 3 for each bosonic variable we introduce a pair of complex conjugate fermionic partners and build novel $N=2$ supersymmetric rational and hyperbolic Ruijsenaars-Schneider three--body models. In contrast to the non--relativistic $N=2$ Calogero models \cite{FM}, the supersymmetry charges involve contributions cubic in the fermionic variables. In the concluding section 4 we discuss possible further developments.

Throughout the paper summation over repeated indices is understood unless otherwise is stated explicitly.

\newpage

\noindent
{\bf 2. Ruijsenaars-Schneider models}\\

\noindent
The Ruijsenaars-Schneider models are integrable many--body systems in one dimension which are described by the equations of motion \cite{RS}
\be\label{RS}
\ddot{x}_i=\sum_{j\ne i} {\dot x}_i {\dot x}_j W(x_i-x_j),
\ee
where $W(x)=\frac{2}{x}$, $\frac{2}{\sinh{x}}$, or $2 \coth{x}$.~\footnote{The so called trigonometric models follow from the hyperbolic systems after the substitution $x\to i x$. In what follows we disregard them.} For simplicity of presentation, in what follows we focus on the three--body problem only and assume $x_1<x_2<x_3$. Note that the models hold invariant under the temporal and spatial translations. The rational system is also invariant under independent rescalings of $t$ and $x_i$ \cite{C1}.

\vspace{0.5cm}
\noindent
{\it 2.1. Rational model }

\vspace{0.5cm}
\noindent
The rational Ruijsenaars-Schneider system corresponds to $W(x)=\frac{2}{x}$ which is also known as the goldfish model \cite{C1}. The equations of motion follow from the Hamiltonian\footnote{The Hamiltonian formulation (\ref{RM}) is not unique \cite{C1}. One can verify that multiplying each term in (\ref{RM}) by an arbitrary constant one does not alter the equations of motion. Keeping in mind the forthcoming construction of an $N=2$ supersymmetric extension, we stick to the Hamiltonian each term of which is positive definite. We also do so for the Hamiltonians in subsect. 2.2 and 2.3.}
\bea\label{RM}
&&
H=\frac{e^{p_1}}{x_{12}x_{13}}+\frac{e^{p_2}}{x_{12}x_{23}}+\frac{e^{p_3}}{x_{13}x_{23}},
\eea
where $x_{ij}=x_i-x_j$ and $(p_1,p_2,p_3)$ signify momenta canonically conjugate to $(x_1,x_2,x_3)$. The Poisson bracket is chosen in the conventional form $\{x_i,p_j \}=\delta_{ij}$.

One of the ways to construct three mutually commuting constants of the motion is to use the Lax matrix \cite{RS,C} which yields
\bea
&&
I_1=H, \quad
I_2=\frac{{\tilde x}_{23} e^{p_1}}{x_{12}x_{13}}+\frac{{\tilde x}_{13} e^{p_2}}{x_{12}x_{23}}+\frac{{\tilde x}_{12} e^{p_3}}{x_{13}x_{23}},
\quad
I_3=\frac{x_2 x_3 e^{p_1}}{x_{12}x_{13}}+\frac{x_1 x_3 e^{p_2}}{x_{12}x_{23}}+\frac{x_1 x_2e^{p_3}}{x_{13}x_{23}},
\eea
where ${\tilde x}_{ij}=x_i+x_j$. These are functionally independent.

The rational model is known to be maximally superintegrable \cite{AF}. Since (\ref{RS}) is translation invariant, the total momentum
\be
I_0=p_1+p_2+p_3
\ee
is conserved. Other constants of the motion are built by considering the elementary monomials
\be\label{Mp}
M_p=\sum_{i_1<\dots<i_p} x_{i_1} \dots x_{i_p}, \qquad \{M_p,H\}=I_p,
\ee
where $p=1,\dots,3$, such that $M_i I_j-M_j I_i$ are conserved quantities. For the case at hand it suffices to
consider
\be
I_4=\frac{x_2 x_3 {\tilde x}_{23}e^{p_1}}{x_{12}x_{13}}+\frac{x_1 x_3 {\tilde x}_{13}e^{p_2}}{x_{12}x_{23}}+\frac{x_1 x_2 {\tilde x}_{12}e^{p_3}}{x_{13}x_{23}}=M_1 I_3-M_3 I_1.
\ee
It is straightforward to verify that $I_k$, $k=0,\dots,4$, are functionally independent which implies the three--body problem (\ref{RM}) is maximally superintegrable.

\vspace{0.5cm}
\noindent
{\it 2.2. Hyperbolic model I }

\vspace{0.5cm}
\noindent
The first of the Ruijsenaars-Schneider hyperbolic models relies upon $W(x)=\frac{2}{\sinh{x}}$. It is described by the Hamiltonian
\bea\label{HM1}
&&
H=e^{p_1} \coth{\left(\frac{x_{12}}{2}\right)}\coth{\left(\frac{x_{13}}{2}\right)}+e^{p_2} \coth{\left(\frac{x_{12}}{2}\right)} \coth{\left(\frac{x_{23}}{2}\right)}
\nonumber\\[2pt]
&&
\qquad
+e^{p_3} \coth{\left(\frac{x_{13}}{2}\right)}\coth{\left(\frac{x_{23}}{2}\right)}=I_1,
\eea
which is chosen such that each term is positive definite (recall $x_1<x_2<x_3$).
Like its rational counterpart (\ref{RM}), the system (\ref{HM1}) is invariant under the spatial translation, $x'_i=x_i+a$, which results in the conservation of the total momentum
\be
I_0=p_1+p_2+p_3.
\ee
The third constant of the motion, which ensures the Liouville integrability, reads
\bea\label{tc}
&&
I_2=e^{p_1+p_2} \coth{\left(\frac{x_{13}}{2}\right)} \coth{\left(\frac{x_{23}}{2}\right)}+e^{p_1+p_3} \coth{\left(\frac{x_{12}}{2}\right)} \coth{\left(\frac{x_{23}}{2}\right)}
\nonumber\\[2pt]
&&
\qquad
+e^{p_2+p_3} \coth{\left(\frac{x_{12}}{2}\right)} \coth{\left(\frac{x_{13}}{2}\right)}.
\eea
One of the ways to obtain (\ref{tc}) is to use the Lax matrix \cite{RS,C}. It is readily verified that $(I_0,I_1,I_2)$ are mutually commuting and functionally independent.

\vspace{0.5cm}

\noindent
{\it 2.3. Hyperbolic model II }

\vspace{0.5cm}

\noindent
The second Ruijsenaars-Schneider hyperbolic model is associated with $W(x)=2 \coth{x}$. We choose the Hamiltonian in the form
\be\label{HM2}
H=\frac{e^{p_1}}{\sinh{x_{12}}\sinh{x_{13}}}+\frac{e^{p_2}}{\sinh{x_{12}}\sinh{x_{23}}}+\frac{e^{p_3}}{\sinh{x_{13}}\sinh{x_{23}}}=I_1.
\ee
Again, in view of $x_1<x_2<x_3$, each term in (\ref{HM2}) is positive definite. Three mutually commuting and functionally independent integrals of motion include (\ref{HM2}) the total momentum
\be
I_0=p_1+p_2+p_3,
\ee
and
\be\label{I2}
I_2=\frac{e^{p_1+p_2}}{\sinh{x_{13}}\sinh{x_{23}}}+\frac{e^{p_1+p_3}}{\sinh{x_{12}}\sinh{x_{23}}}+\frac{e^{p_2+p_3}}{\sinh{x_{12}}\sinh{x_{13}}}.
\ee
The simplest way to obtain (\ref{I2}) is to use the Lax matrix \cite{RS,C}.

\vspace{0.5cm}
\noindent
{\it 2.4. Geodesic interpretation}

\vspace{0.5cm}
\noindent
The Ruijsenaars-Schneider equations of motion (\ref{RS}) can be rewritten as the geodesic equations on a manifold which is parametrized by the local coordinates $x_i$ and equipped with the affine connection (no summation over repeated indices) \cite{ABHL}
\bea\label{CS}
\Gamma^i_{jk}=\delta^i_j w_{ik}+\delta^i_k w_{ij}, \qquad w_{ik}=\left\{
\begin{aligned}
-\frac 12 W(x_i-x_k),&\qquad i\ne k\\
0 \qquad \qquad ,&\qquad i=k\\
\end{aligned}
\right.
\eea
For the rational model (\ref{CS}) turns out to be a metric connection associated with \cite{ABHL}
\be\label{Metr}
g_{ij}=\frac{\partial M_p}{\partial x_i} \frac{\partial M_p}{\partial x_j},
\ee
where the functions $M_p$ are given in (\ref{Mp}) with $p=1,\dots, n$. Since (\ref{Metr}) is the Kronecker delta in curvilinear coordinates, the transformation $x'_i=M_i(x)$ links the rational Ruijsenaars-Schneider model to a free particle propagating in a flat space.

Let us examine whether the hyperbolic choices of $W(x)$ result in metric connections. Assuming a metric is  non--degenerate and (\ref{CS}) can be represented in the conventional form
\be
\Gamma^i_{jk}=\frac 12 g^{ip} \left(\partial_j g_{pk}+\partial_k g_{pj}-\partial_p g_{jk} \right),
\ee
contracting with $g_{si}$, permuting the indices $(j,s,k)\to(s,k,j)$, and taking the sum, one gets a coupled set of partial differential equations
\be\label{sys}
\partial_j g_{sk}=w_{jk}(g_{sj}-g_{sk})+w_{js}(g_{kj}-g_{ks}).
\ee
It turns out that (\ref{sys}) leads to a contradiction as it yields a degenerate metric whose all components are equal to one and the same constant, $g_{ij}=\mbox{const}$.
In order to see this, it suffices to consider three equations belonging to the set (\ref{sys})
\be
\partial_1 g_{11}=0, \qquad \partial_2 g_{11}=2 w_{12}(g_{11}-g_{12}), \qquad \partial_1 g_{12}=w_{12} (g_{11}-g_{12}).
\ee
\begin{table}
\caption{Functions $\lambda_i$ for the Ruijsenaars-Schneider models}\label{table}
\begin{eqnarray*}
\begin{array}{|c|c|c|c}
\hline
W(x)=\frac{2}{x} & W(x)=\frac{2}{\sinh{x}} & W(x)=2 \coth{x}
\\
\hline
\lambda_1=\frac{e^{\frac{p_1}{2}}}{\sqrt{x_{12} x_{13}}} & \lambda_1=e^{\frac{p_1}{2}} \sqrt{ \coth{\left(\frac{x_{12}}{2}\right)}\coth{\left(\frac{x_{13}}{2}\right)}} & \lambda_1=\frac{e^{\frac{p_1}{2}}}{ \sqrt{ \sinh{x_{12}} \sinh{x_{13}}}} \\
\hline
\lambda_2=\frac{e^{\frac{p_2}{2}}}{\sqrt{x_{12} x_{23}}} & \lambda_2=e^{\frac{p_2}{2}}\sqrt{ \coth{\left(\frac{x_{12}}{2}\right)} \coth{\left(\frac{x_{23}}{2}\right)}} & \lambda_2=\frac{e^{\frac{p_2}{2} }}{ \sqrt{ \sinh{x_{12}} \sinh{x_{23}}}} \\
\hline
\lambda_3=\frac{e^{\frac{p_3}{2}}}{\sqrt{x_{13} x_{23}}} & \lambda_3=e^{\frac{p_3}{2} } \sqrt{\coth{\left(\frac{x_{13}}{2}\right)}\coth{\left(\frac{x_{23}}{2}\right)}} & \lambda_3=\frac{e^{\frac{p_3}{2} }}{ \sqrt{ \sinh{x_{13}} \sinh{x_{23}}}} \\
\hline
\end{array}
\end{eqnarray*}
\end{table}

Computing the derivative of the second equation with respect to $x_1$ and taking into account the other two, one gets
\be
\left(w'_{12}-w_{12}^2 \right)(g_{11}-g_{12})=0,
\ee
where $w'=\frac{d w(x)}{dx}$. Since for the hyperbolic models $\left(w'_{12}-w_{12}^2 \right)\ne 0$, one obtains
\be
g_{11}=g_{12}.
\ee
By repeatedly using the same argument, one can further demonstrate that all components of $g_{ij}$ are equal to each other. The left hand side of (\ref{sys}) then implies $g_{ij}=\mbox{const}$.

Thus, in contrast to the rational model, the hyperbolic Ruijsenaars-Schneider systems are linked to non--metric connections. While in the former case all components of the Riemann tensor vanish identically, in the latter case the curvature tensor is non--trivial.

\vspace{0.5cm}

\noindent
{\bf 3. $N=2$ supersymmetric extension of Ruijsenaars-Schneider models}\\

\noindent
As was emphasized above, the Hamiltonian formulations for the Ruijsenaars-Schneider models were chosen so that each term in the  Hamiltonian was positive definite.
In order to construct $N=2$ supersymetric extensions, we first represent the original bosonic Hamiltonian in the form
\be
H_B=\lambda_i \lambda_i,
\ee
where the phase space functions $\lambda_i$, $i=1,2,3$, are given above in Tab. 1. They prove to obey the quadratic algebra (no summation over repeated indices and $i\ne j$)
\be\label{All}
\{\lambda_i,\lambda_j\}=\frac 14 W(x_i-x_j) \lambda_i \lambda_j.
\ee
Note that this algebra holds invariant under the rescalings $\lambda_i \to \alpha_i \lambda_i$ (no sum), where $\alpha_i$ are arbitrary real constants. This transformation links to the arbitrariness in the choice of the Hamiltonian mentioned above.

Then we introduce the complex fermionic partners $\psi_i$, $i=1,2,3$, for the bosonic coordinates $x_i$, and impose the canonical brackets
\be
\{\psi_i,\psi_j\}=0, \qquad \{\psi_i,\bar\psi_j\}=-i \delta_{ij}, \qquad  \{\bar\psi_i,\bar\psi_j\}=0,
\ee
where $\bar\psi_i$ stands for the complex conjugate of $\psi_i$.

Two supersymmetry charges are chosen in the polynomial form
\be\label{SUSY}
Q=\lambda_i \psi_i+i f_{ijk} \psi_i \psi_j {\bar\psi}_k, \qquad  \bar Q=\lambda_i \bar\psi_i+i f_{ijk} \bar\psi_i \bar\psi_j \psi_k,
\ee
where $f_{ijk}=-f_{jik}$ are real functions. The latter are determined from the condition that the supersymmetry charge is nilpotent $\{Q,Q\}=0$:
\bea\label{EF}
&&
\{\lambda_i,\lambda_j \}+2 f_{ijk} \lambda_k=0, \qquad \{ \lambda_{\underline{k}} ,f_{\underline{n} \underline{m} l} \}+2 f_{\underline{k} \underline{n} p} f_{p \underline{m} l}=0, \qquad \{f_{{\underline{a}} \underline{b} \overline{c}}, f_{\underline{m} \underline{n} \overline{k}} \}=0,
\eea
where the underline/overline mark signifies antisymmetrization of the respective indices.

The Hamiltonian which governs the dynamics of an $N=2$ supersymmetric extension follows from the superalgebra
\be
\{Q,\bar Q \}=-i H,
\ee
which yields
\bea\label{HamM}
&&
H=\lambda_i \lambda_i-2i (f_{ijk}+f_{kji}+f_{ikj})\lambda_k \psi_i \bar\psi_j+i\{f_{ijl},f_{mnk} \}\psi_i \psi_j \psi_k \bar\psi_l \bar\psi_m \bar\psi_n
\nonumber\\[2pt]
&&
\qquad -(\{\lambda_i,f_{klj}\}-\{\lambda_l,f_{ijk}\}+f_{ijp}f_{klp}-4f_{pil}f_{pkj})\psi_i \psi_j \bar\psi_k \bar\psi_l.
\eea

Comparing (\ref{All}) with the leftmost equation in (\ref{EF}), one gets
\begin{align}\label{ff}
&
f_{121}=-\frac{a}{8} W(x_1-x_2)\lambda_2, && f_{122}=-\frac{(1-a)}{8} W(x_1-x_2) \lambda_1,
\nonumber\\[2pt]
& f_{131}=-\frac{b}{8} W(x_1-x_3)\lambda_3, &&
f_{133}=-\frac{(1-b)}{8} W(x_1-x_3) \lambda_1,
\nonumber\\[2pt]
& f_{232}=-\frac{c}{8} W(x_2-x_3)\lambda_3, && f_{233}=-\frac{(1-c)}{8} W(x_2-x_3)\lambda_2,
\end{align}
where $(a,b,c)$ are arbitrary real constants, while other components of $f_{ijk}$ prove to vanish. Substituting (\ref{ff}) into the second equation in (\ref{EF}), one obtains the quadratic algebraic equations
\be\label{str}
bc=0, \qquad a(1-c)=0, \qquad (1-a)(1-b)=0,
\ee
which imply that two options are available
\be\label{str1}
a=1, \qquad b=0, \qquad c=1,
\ee
or
\be\label{str2}
a=0, \qquad b=1, \qquad c=0.
\ee
It is straightforward to verify that the second possibility is linked to the first by relabelling
\be
x_1 \leftrightarrow x_3, \qquad p_1 \leftrightarrow p_3, \qquad \psi_1 \leftrightarrow \psi_3, \qquad \bar\psi_1 \leftrightarrow \bar\psi_3,
\ee
which gives $\lambda_1 \leftrightarrow \lambda_3$, $\lambda_2 \leftrightarrow \lambda_2$. For the three--body problem the rightmost equation in (\ref{EF}) holds automatically.

Thus, $N=2$ supersymmetric extensions of the Ruijsenaars-Schneider models build upon $\lambda_i$, which are exposed above in Tab. 1, and the structure functions
\be
f_{121}=-\frac{1}{8} W(x_1-x_2)\lambda_2, \qquad  f_{133}=-\frac{1}{8} W(x_1-x_3) \lambda_1, \qquad f_{232}=-\frac{1}{8} W(x_2-x_3)\lambda_3.
\ee
Interestingly enough, in contrast to $N=2$ supersymmetric extensions of the non--relativistic Calogero model \cite{FM}, the supersymmetry charges involve contributions cubic in the fermionic variables. Thus, provided one is focused on a Hamiltonian each term of which is positive definite, the $N=2$ supersymmetric extension is essentially unique.

It proves instructive to expose the (complex) supersymmetry charge and the Hamiltonian in terms of $\lambda_i$ and the prepotential $W(x)$
\bea
&&
Q=\lambda_1 \psi_1+\lambda_2 \psi_2+\lambda_3 \psi_3-\frac{i}{4} W(x_1-x_2) \lambda_2 \psi_1 \psi_2 \bar\psi_1-\frac{i}{4} W(x_1-x_3) \lambda_1 \psi_1 \psi_3 \bar\psi_3
\nonumber\\[2pt]
&& \quad \quad
-\frac{i}{4} W(x_2-x_3) \lambda_3 \psi_2 \psi_3 \bar\psi_2,
\nonumber\\[2pt]
&&
H=\lambda_1^2+\lambda_2^2+\lambda_3^2+\frac{i}{2} W(x_1-x_2) \lambda_1 \lambda_2 (\psi_1 \bar\psi_2 -\psi_2 \bar\psi_1)+\frac{i}{2} W(x_1-x_3) \lambda_1 \lambda_3 (\psi_1 \bar\psi_3 -\psi_3 \bar\psi_1)
\nonumber\\[2pt]
&&
\quad \quad +\frac{i}{2} W(x_2-x_3) \lambda_2 \lambda_3 (\psi_2 \bar\psi_3 -\psi_3 \bar\psi_2)-\frac 14 W'(x_1-x_2)\lambda_2^2 \psi_1 \psi_2 \bar\psi_1\bar\psi_2
\nonumber\\[2pt]
&&
\quad \quad
-\frac 14 W'(x_1-x_3)\lambda_1^2 \psi_1 \psi_3 \bar\psi_1\bar\psi_3-\frac 14 W'(x_2-x_3)\lambda_3^2 \psi_2 \psi_3 \bar\psi_2\bar\psi_3
\nonumber\\[2pt]
&&
\quad \quad
+\frac 18 W(x_1-x_2) W(x_1-x_3) \lambda_1 \lambda_2 \psi_3 \bar\psi_3 (\psi_1 \bar\psi_2 +\psi_2 \bar\psi_1)
\nonumber\\[2pt]
&&
\quad \quad
+\frac 18 W(x_1-x_3) W(x_2-x_3) \lambda_1 \lambda_3 \psi_2 \bar\psi_2 (\psi_1 \bar\psi_3 +\psi_3 \bar\psi_1)
\nonumber\\[2pt]
&&
\quad \quad
-\frac 18 W(x_1-x_2) W(x_2-x_3) \lambda_2 \lambda_3 \psi_1 \bar\psi_1 (\psi_2 \bar\psi_3 +\psi_3 \bar\psi_2),
\eea
where $W'(x)=\frac{dW(x)}{dx}$.
Curiously enough, for the three--body models the six--fermion term present in (\ref{HamM}) proves to be zero. We failed to demonstrate that it also vanishes for $n>3$ on account of Eqs. (\ref{EF}).

\vspace{0.5cm}

\noindent
{\bf 4. Conclusion}\\

\noindent
The construction of the $N=2$ supersymmetric rational and hyperbolic Ruijsenaars-Schneider three--body models reported in this work can be continued in several directions.

First of all, it is worth extending the present analysis to the case of arbitrary number of particles. For the rational model an optimal strategy might be to switch to the geodesic formulation associated with the metric (\ref{Metr}). One can first implement a coordinate transformation which brings the model to the free form, supersymmetrize the free system, and then apply the inverse transformation. A canonical transformation linking such a system to (\ref{RM}) for $n=3$ is of interest. For the hyperbolic models the construction may break beyond $n=3$. For the case of $n$ particles the structure functions $f_{ijk}$ involve $n C_n^2$ components, where $C_m^k$ are the binomial coefficients. The first, second, and third equations in (\ref{EF}) yield $C_n^2$, $n C_n^3$, and $C_n^2 C_n^4$ conditions, respectively. For $n>3$ the set of restrictions is overcomplete. In particular, some of them may turn out to be incompatible with the form of the prepotential $W(x)$ chosen.

Secondly, it is interesting to construct an off--shell superfield Lagrangian formulation for the on--shell component Hamiltonian (\ref{HamM}) and to study its peculiarities.

Thirdly, an $N=4$ supersymmetric generalization is an intriguing open problem. The key point is to reveal an analogue of the Witten-Dijkgraaf-Verlinde-Verlinde equation \cite{GLP}. As was mentioned above, the hyperbolic Ruijsenaars-Schneider models can be described in terms of the geodesic equations associated with a non--metric connection. The description of many--body mechanics with extended supersymmetry on such spacetimes in purely geometric terms is a challenge.

Finally, it would be interesting to understand whether supersymmetric extensions of the Ruijsenaars-Schneider models may be relevant for the study of the space  of  vacua  of supersymmetric gauge theories (see the discussion in \cite{GK} and references therein).

\vspace{0.5cm}

\noindent{\bf Acknowledgements}\\

\noindent
This work was supported by the Tomsk Polytechnic University competitiveness enhancement program.

\end{document}